\documentclass[twoside,slac_one]{revtex4}
\usepackage{graphicx}
\usepackage{fancyhdr}
\usepackage{amsmath}
\usepackage{bm}
\usepackage{amsxtra}
\usepackage{amssymb}
\usepackage{amsthm}
\usepackage{latexsym}
\usepackage{lscape}

\RequirePackage{xspace}

\def\invpb {\ensuremath{\mbox{\,pb}^{-1}}\xspace}

\def\invfb   {\ensuremath{\mbox{\,fb}^{-1}}\xspace}
\newcommand{\tev}{\ensuremath{\mathrm{\,Te\kern -0.1em V}}\xspace}
\newcommand{\gev}{\ensuremath{\mathrm{\,Ge\kern -0.1em V}}\xspace}
\newcommand{\mev}{\ensuremath{\mathrm{\,Me\kern -0.1em V}}\xspace}
\newcommand{\kev}{\ensuremath{\mathrm{\,ke\kern -0.1em V}}\xspace}
\newcommand{\ev}{\ensuremath{\mathrm{\,e\kern -0.1em V}}\xspace}
\def\pt {\ensuremath{p_T}\xspace}
\def\et {\ensuremath{E_T}\xspace}
\def\cm   {\ensuremath{\rm \,cm}\xspace}
\def\beq {\begin{equation}}
\def\eeq {\end{equation}}
\def\Zpr      {\ensuremath{{Z^0}^\prime}\xspace}

\def\ppzll  {\ensuremath{\proton\proton\to\Z\to\ellell}\xspace}

\def\zll  {\ensuremath{\Z\to\ellell}\xspace}
\def\zprll  {\ensuremath{\Zpr\to\ellell}\xspace}
\def\selEff{\ensuremath{\epsilon_\mathtt{sel}}\xspace}
\def\acc{\ensuremath{{\cal A}}\xspace}
\def\xsecz{\ensuremath{\sigma_{\zll}}\xspace}
\def\xseczp{\ensuremath{\sigma_{\zprll}}\xspace}

\def\bw{\ensuremath{\mathtt{BW}}\xspace}
\def\gaus{\ensuremath{\mathtt{Gaussian}}\xspace}

\def\llr {\ensuremath{\lambda}\xspace}
\def\btheta {\ensuremath{{\boldsymbol\theta}}\xspace}
\def\bnu {\ensuremath{{\boldsymbol\nu}}\xspace}
\def\bx {\ensuremath{{\boldsymbol x}}\xspace}
\def\bm {\ensuremath{{\boldsymbol m}}\xspace}
\newcommand{\chisq}{\ensuremath{\chi^2}\xspace}

\def\ellell     {\ensuremath{\ell^+ \ell^-}\xspace}
\def\Z      {\ensuremath{Z^0}\xspace}
\def\proton      {\ensuremath{p}\xspace}
\def\m    {\ensuremath{\rm \,m}\xspace}
\def\mll{\ensuremath{m_{ll}}\xspace}

\begin{document}

\title{Search for narrow resonances in the lepton final state at CMS}

\author{G. Kukartsev}
\affiliation{Department of Physics and Astronomy, Brown University, Providence, RI, USA}

\begin{abstract}
We discuss the results of searches for high-mass narrow resonances decaying 
into pairs of leptons using pp collisions at 7 TeV delivered by 
LHC and collected with the CMS detector in 2010 and 2011. These 
include searches for the \Zpr bosons and RS gravitons.
\end{abstract}

\maketitle

\thispagestyle{fancy}

\section{Introduction}
Several theoretical models predict new \tev-scale resonances 
decaying into a pair of leptons. Models of particular interest
for the presented analysis 
include the Sequential Standard Model (SSM) with standard-model-like
couplings, and certain grand-unification-motivated 
models ($\Psi$) \cite{Leike:1998wr}. Both predict narrow \Z-boson-like
states (\Zpr). We also consider Kaluza-Klein excitations in
the Randall-Sundrum (RS) model of extra 
dimensions ($G_\mathtt{KK}$) \cite{Randall:1999vf,Randall:1999ee}. We use the
four listed models as benchmarks while we search for a narrow
resonance, which is
similar to the SSM \Zpr,
in the dimuon 
and the dielectron channels.
We perform
a likelihood-based shape analysis of the reconstructed dilepton 
invariant mass (\mll) spectra. The approach provides robustness
against uncertainties in the absolute background rate.

The recent searches for $\Zpr\to l^{+}l^{-}$ and 
$G_\mathtt{KK}\to l^{+}l^{-}$ were published by the Tevatron 
experiments \cite{D0_RS,D0_Zp,CDF_RS,CDF_Zp}.
There are indirect constraints from
LEP-II \cite{delphi,aleph,opal,l3}.

\section{Detector and Experiment}
\begin{figure*}[ht]
\centering
\includegraphics[width=135mm]{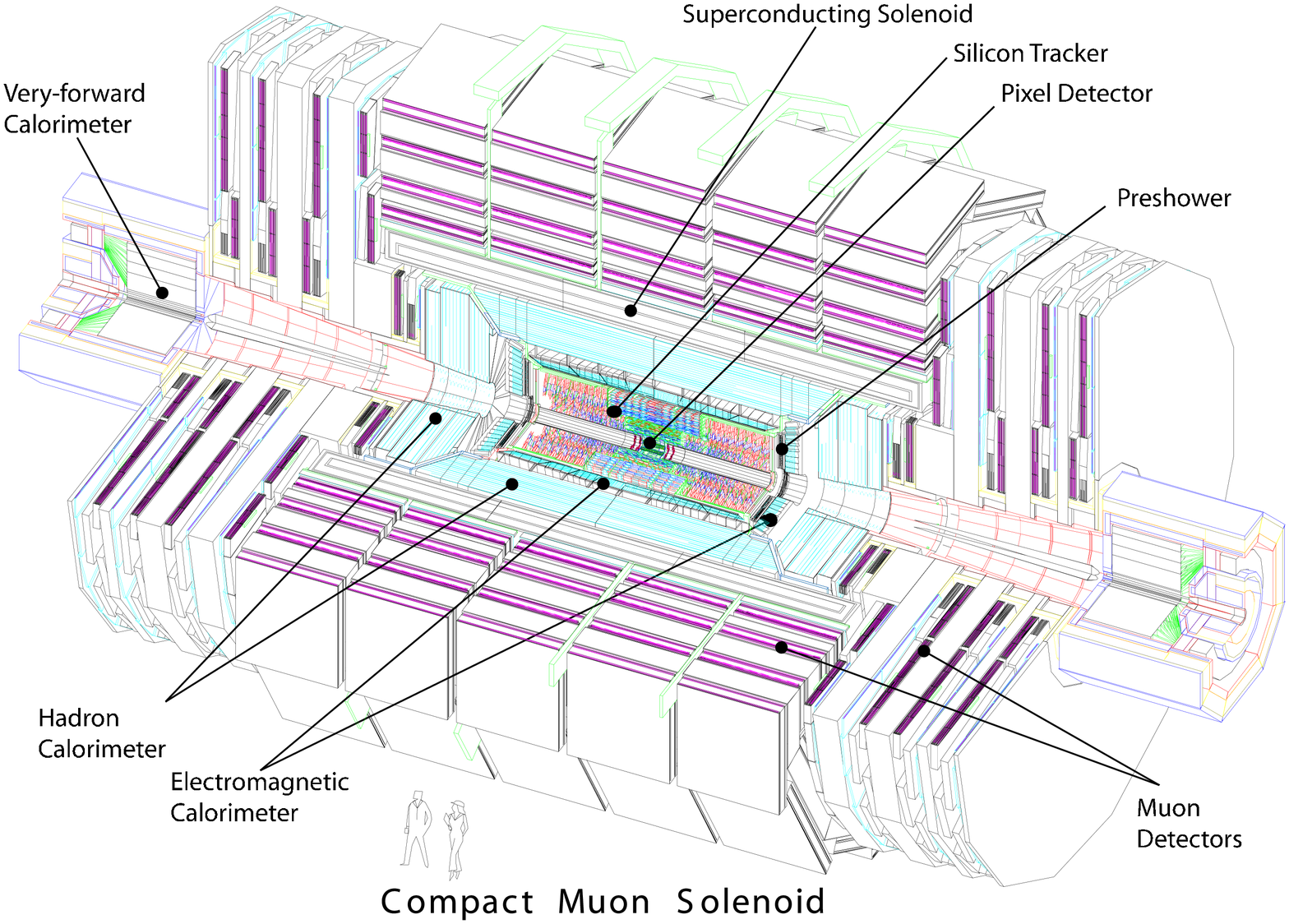}
\caption{The CMS detector.} \label{fig:cms}
\end{figure*}

CMS is a general-purpose particle detector located
at the LHC proton-proton collider at CERN.
A prominent feature of the detector is
a superconducting solenoid with the internal diameter of 6\m
and an axial field of 3.8~T. The solenoid encloses 
the pixel detector, the silicon tracker, 
the crystal electromagnetic calorimeter (ECAL)
and the brass and scintillator hadron calorimeter (HCAL).
Outside the solenoid there is a steel flux return yoke
instrumented with the gas ionization detectors, which
constitute the CMS muon system.
A diagram of the detector is shown in Figure~\ref{fig:cms}.
Further details can be found elsewhere \cite{JINST}. For the
presented results, 1.1\invfb of integrated luminosity were
used.

\section{Data and Monte Carlo}
The presented results were obtained using the data recorded
by the CMS experiment in 2011. The data were taken using
proton-proton colliding beams with the center-of-mass energy of 
$7\tev$. The size of the dataset corresponds
to an integrated luminosity of approximately $1.1\invfb$. The size of
the data sample used in the dielectron analysis is $25\invpb$
smaller due to different quality requirements for the data.

The signal and background processes were modeled using
Monte Carlo simulations. Depending on the process,
\texttt{PYTHIA v6.424} \cite{Sjostrand:2006za},
\texttt{MADGRAPH} \cite{MADGRAPH}
and \texttt{POWHEG v1.1} \cite{Alioli:2008gx,Nason:2004rx,Frixione:2007vw}
event generators
together with 
the \texttt{CTEQ6L1} \cite{Pumplin:2002vw}
parton distribution function (PDF) set
were used.
The full CMS detector simulation was done with
\texttt{GEANT4} \cite{GEANT4}.
The generated events were passed through
the CMS trigger simulation and full reconstruction sequence.

\section{Event Selection}
We developed dedicated selection criteria for each of the two
dilepton channels under consideration. Even though the underlying
physics processes under study are similar, reconstruction of
different lepton flavors in the detector differs substantially.
For electrons, we reconstruct the transverse energy using
calorimeter information, while the muon reconstruction is 
based on
the tracking and the muon systems for the measurement
of the transverse momenta.
Dilepton
invariant mass reconstruction deteriorates for higher values
in the dimuon channel and improves in the dielectron channel.
We require the muons to be reconstructed with the opposite
charge, and do not impose this restriction on dielectron pairs.
For the dimuon pairs reconstruction,
we reduce systematic uncertainty
by performing data-driven studies with cosmic-ray muons. 
The dielectron channel entails higher background rates
from misreconstructed strong scattering signal, and requires
tighter selection, which leads to lower efficiency and acceptance.

\subsection{Trigger}
For the dimuon pair event candidates, we used a single muon trigger
with sufficiently high minimum transverse momentum 
requirement ($\pt>30\gev$). The
muon was firstly required to be detected in the muon system, and
then matched to a track in the silicon tracker. For dielectron
pairs, the trigger requires two sufficiently energetic deposits
($33\gev$) in ECAL, with at
least one of the deposits matched to level-one deposit.
The corresponding deposit in HCAL must be small
(less than $15\%$). In later portions of the dataset, a match to
the activity in the silicon pixel tracker was required.

\subsection{Lepton Reconstruction and Pile-up}
Standard CMS techniques apply to the reconstruction, 
calibration and identification of 
the leptons \cite{MUO-10-004-PAS,EGMPAS,EWK-10-002-PAS}.
For all leptons, the reconstructed track was required to be
consistent with the beam interaction 
point,
to be topologically
isolated from the hadronic signatures, and to be sufficiently
energetic in the plane transverse to the beam 
axis ($\pt>35\gev$ for muons and electrons in the ECAL barrel,
$\pt>40\gev$ for endcap electrons). The muons
are then reconstructed via a global fit of the tracker and the
muon system information with proper quality requirements met:
there should be enough hits (more than 10) in the silicon tracker,
at least 1 hit in the pixel detector,
and a track reconstructed in the tracker and extrapolated to
the muon system must be compatible with the hits in the muon
system, with hits in at least 2 of the muon stations.
The transverse impact parameter relative to the beam interaction
point is required to be less than $0.2\cm$.
The electrons are reconstructed as an ECAL cluster matched
to a track in the silicon tracker. The ECAL cluster seeds the
track in the pixel detector, which in turn seeds the track
in the tracker. Each track must have at least five hits, and a hit
in each of the three pixel layers. The reconstructed electron
candidate must be within either barrel ($|\eta|<1.442$) 
or endcap ($1.56<\eta<2.5$) ECAL acceptance
regions, and less than $5\%$ of the energy must be deposited in 
HCAL.

Leptons are required to be isolated from other activity in
the tracker, in order to suppress background from jets 
misreconstructed as leptons, and from non-prompt leptons.
The isolation is defined using a cone
$\delta R=\sqrt{(\delta\eta)^2+(\delta\phi)^2}$
centered on the lepton
axis
where $\eta$ is pseudo-rapidity
and $\phi$ is the azimuthal angle relative to the beam axis.

For the muon, the sum of transverse momenta of all other tracks,
consistent with the primary vertex,
in the cone of $0.3$ must be less than $10\%$ of the muon \pt.
The efficiency of this isolation requirement was shown to be 
stable with the number of primary vertexes as indication of
robustness against pile-up in a higher instantaneous luminosity
regime.

For the electron, the sum of all track \pt in the cone of $0.04$
is required to be less than $7\gev$
in the barrel
of ECAL, and less than $15\gev$ for the endcap. The tracks are 
required to be consistent with the reconstructed primary vertex.
The calorimeter isolation for the electrons requires that the
sum of \et of all deposits in the ECAL and 
the HCAL to be less than 
$0.03\et+2\gev$ relative to the the electron \et.
For the electrons in endcap, we exploit the HCAL
segmentation along the beam axis. The isolation
energy is required to be less than
$0.03 \cdot \max{(0,\et-50\gev)}+2.5\gev$
where \et is determined from ECAL and the first layer of HCAL.
In the second layer, the HCAL \et must be less than $0.5\gev$.
Additionally, the shape of the transverse energy deposit is 
required to be compatible with the expected electron signal,
and a good match in $\eta$ and $\phi$ with the corresponding
track is required.

\subsection{Lepton pair selection}
We select events with two reconstructed leptons: either muons 
or electrons, originating from a well-reconstructed primary
vertex. The vertex must be within $2\cm$ from the beam interaction
point in the transverse plane, and within $24\cm$ along the
beam axis, to suppress cosmic ray background. For the muon pair
event candidates, an additional protection against cosmic muons
is required as an opening angle between the two muons being
less than $(\pi-0.02)$.

For the dimuon events, we require opposite charges for the
two muons 
as it reduces the fraction of events with 
a large mismeasurement of the
momentum. We suppress events with many poorly
reconstructed tracks in order to reduce beam background.
At least one muon has to match a high-level trigger (HLT)
muon. As an additional quality requirement, the muon pair
is required to be consistent with a common vertex.

For the electron pair events, at least one of the electrons
is required to be reconstructed in the barrel part of the
detector. In order to suppress background from photon conversions,
we impose requirement on the distance to the nearest track and
an opening angle with it.

\subsection{Efficiency and Acceptance}
\begin{figure*}[ht]
\centering
\includegraphics[width=80mm]{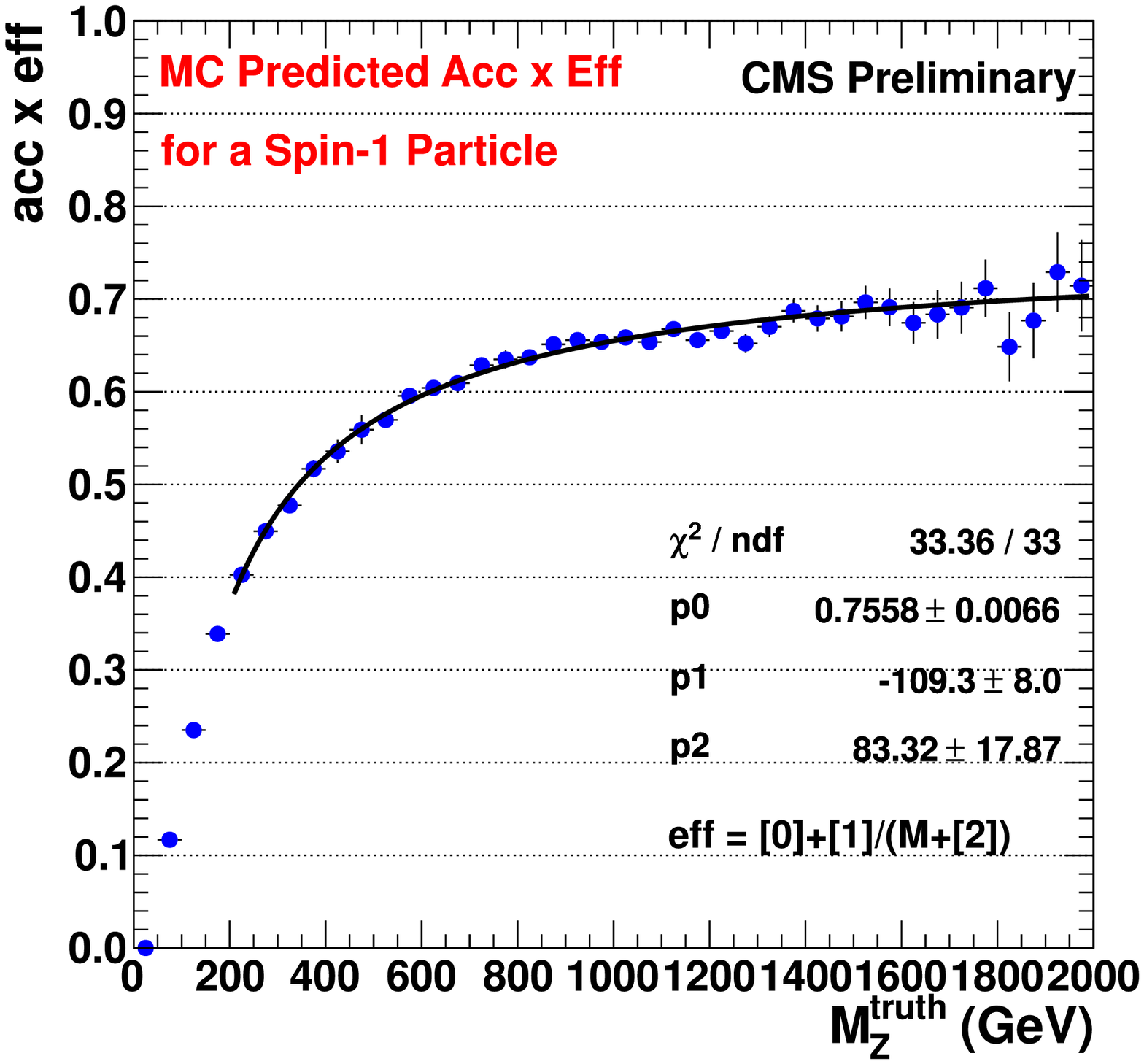}
\includegraphics[width=80mm,height=74mm]{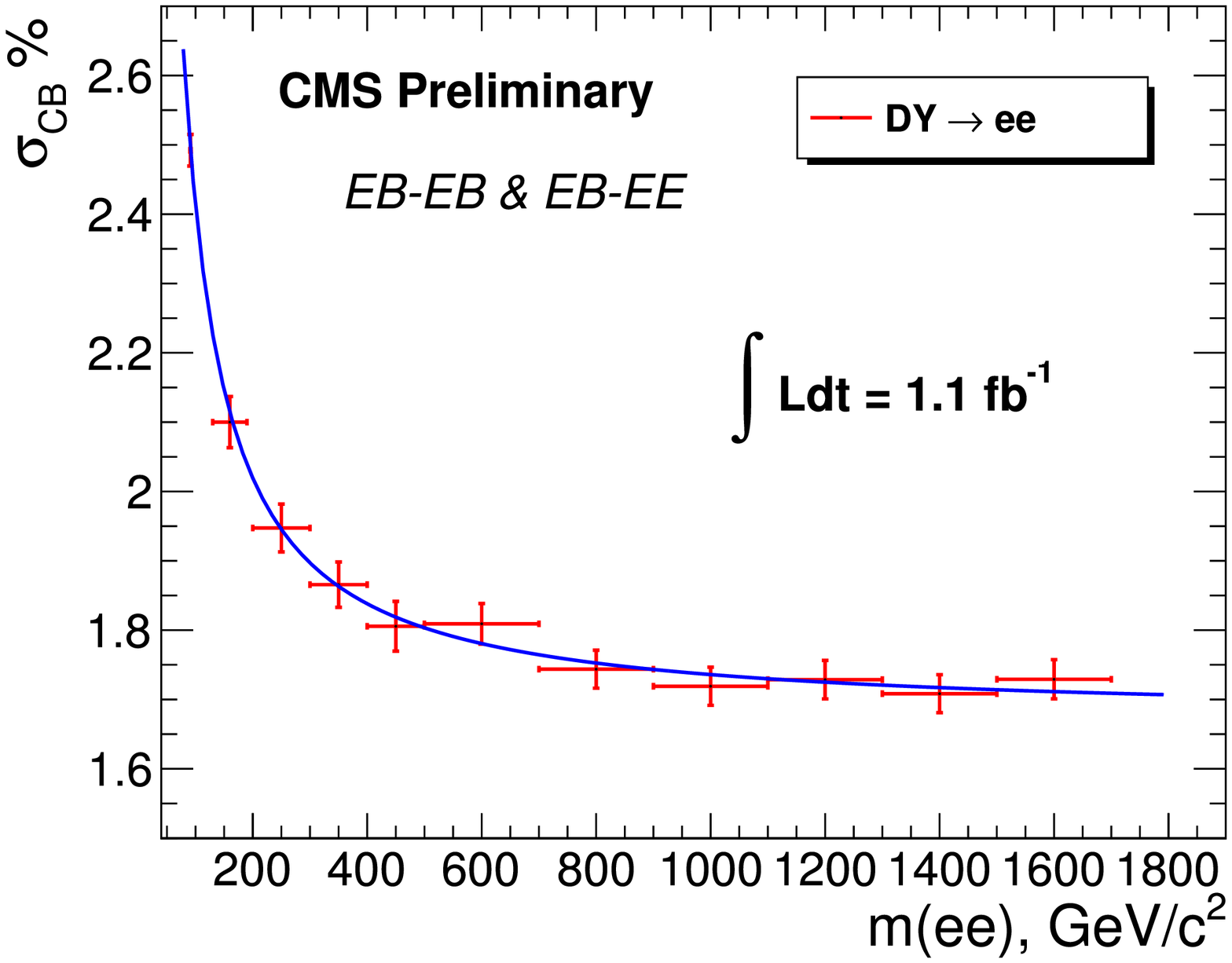}
\caption{Acceptance and efficiency (left)
  and  
  invariant mass  resolution (right), dielectron channel. }
\label{fig:res}
\label{fig:acceff}
\end{figure*}

We measure efficiency of triggering, lepton reconstruction
and identification with ``tag-and-probe'' 
method \cite{MUO-10-004-PAS,EWK-10-002-PAS}. We use a pure
sample of dimuon pairs requiring that their invariant mass is
consistent with the Z boson mass ($60\gev<m_\mathtt{ll}<120\gev$).
One of the muons in the pair is reconstructed with stringent
quality requirements (tag), and the other is used as a probe
for efficiency estimates. Contributing factors also include
track reconstruction and electron clustering.
We measure the single muon trigger efficiency to be
$95.0\%\pm0.3\%$ in the barrel and $89.9\%\pm0.4\%$ in the endcap.
The efficiency of the muon identification is measured to be
$96.4\%\pm0.2\%$ in the barrel and $96.0\%\pm0.3\%$ in the endcap.
The efficiency of the track reconstruction in the internal tracker
is found to be above $99\%$ in the whole acceptance range.
Figure~\ref{fig:acceff} represents the overall acceptance and
efficiency values for the dielectron channel,
as a function of the dilepton invariant mass.
Similar behavior with higher overall acceptance and efficiency
values is observed in the dimuon channel.

\section{Resolution}
We study detector performance using Standard Model processes with
W and Z mesons and their leptonic final states. We also use cosmic
muons.
The muon momentum resolution ranges from $1\%$ at few tens of \gev
(Z boson peak scale) to approximately $10\%$ above $1\tev$.
Tracker-based reconstruction yields better performance at low
momenta, while the muons reconstructed in the muon system have
better resolution at high momenta. However, energy loss in the
steel yoke and showers in the muon chambers can spoil the global
fit. We find that adding muon
system hits to the tracker-based fit improves resolution
for muons with \pt greater than approximately
$200\gev$ \cite{PTDR2}. 
The most comprehensive algorithm ("Tune P'') 
makes track-by-track decisions about which hits in which
subsystems to use. The resolution is also sensitive to the 
alignment of the muon and the tracker systems.

Unlike for muons, the electron energy resolution improves with
energy. The ECAL resolution is better
than $0.5\%$ for unconverted photons with transverse energies
above $100\gev$. The invariant mass resolution of
dielectron pairs is modeled with a Crystal Ball function and
obtained from Monte Carlo simulation,
with additional smearing applied. The smearing is obtained from
comparisons of the Z-boson peak fits in data and Monte Carlo
simulation of the $Z\to ee$ process.
At $1\tev$, the dielectron invariant mass resolution is
approximately $1.3\%$ when both electrons are in the barrel
acceptance region, and approximately $2.4\%$ when one of
the electrons is in the endcap region. For the electrons in the
barrel section of the detector, energy scale is established
using neutral pions and checked using the Z peak.

\section{Background\label{sec:bg}}
\begin{figure*}[ht]
\centering
\includegraphics[width=80mm]{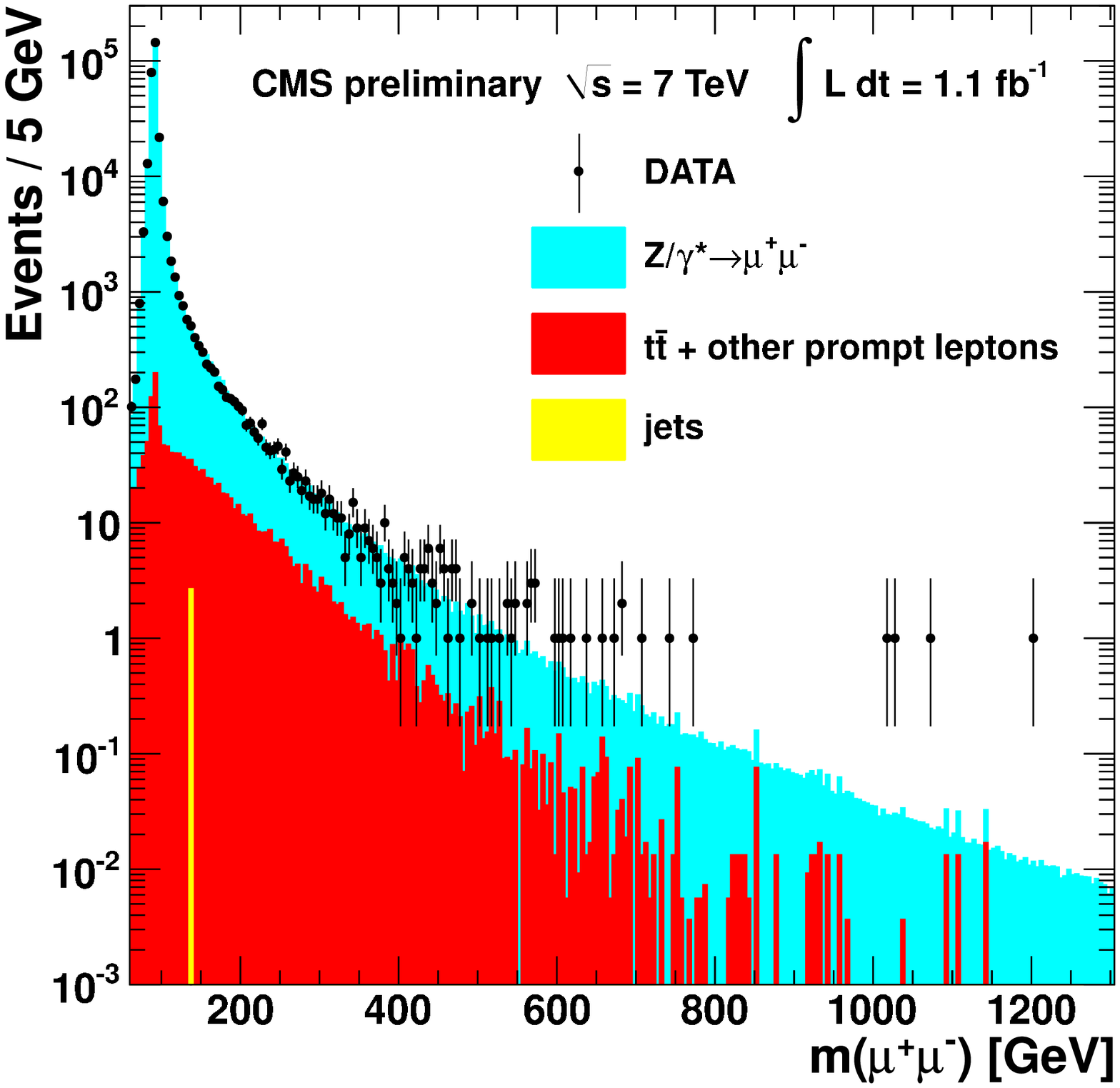}
\includegraphics[width=80mm]{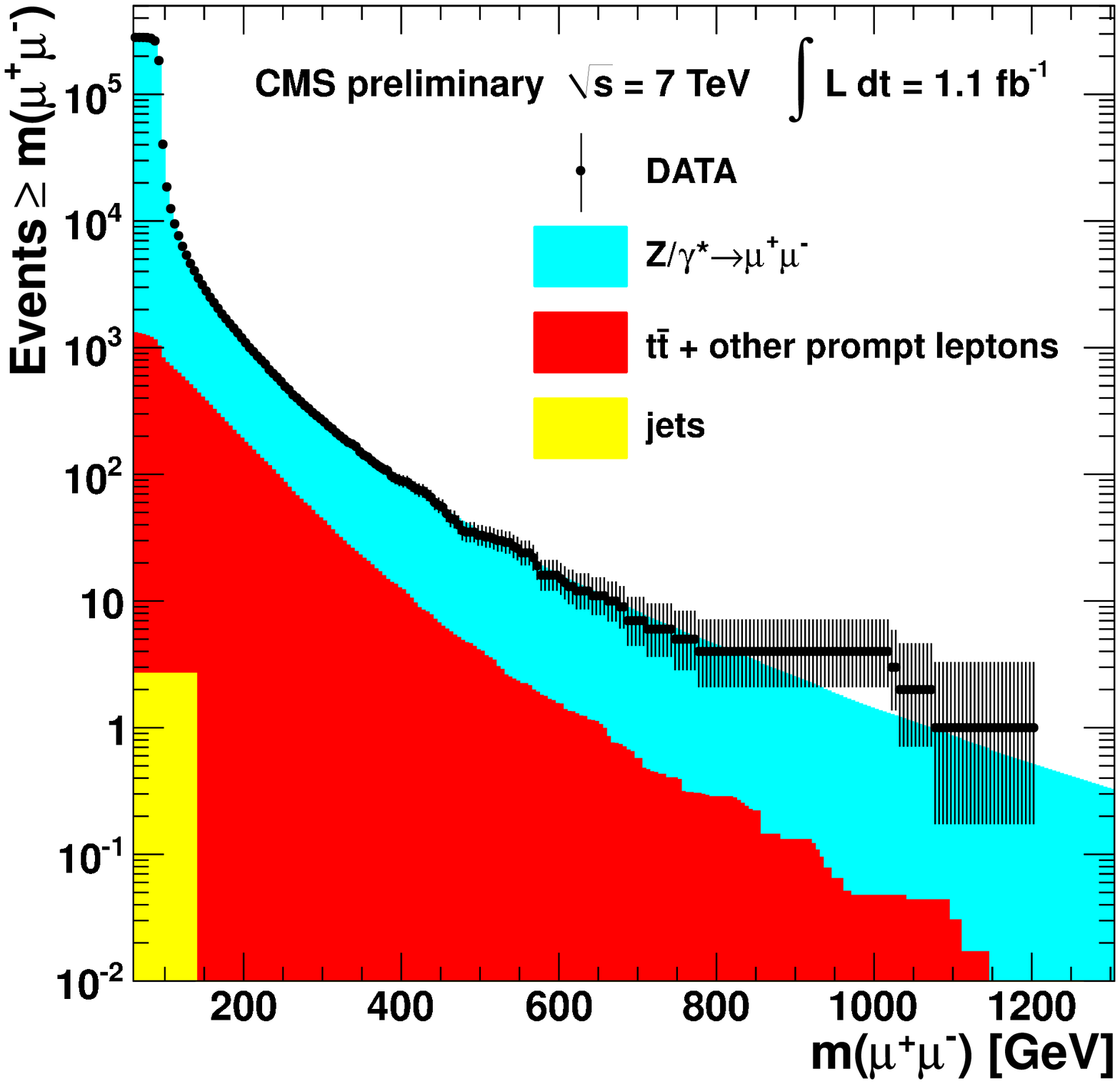}
\caption{Dimuon invariant mass (left) and the corresponding 
  cumulative spectrum (right).
  Individual components are normalized to NLO and then together
  to the Z-boson peak.
}
\label{fig:dimuon_mass}
\end{figure*}

\begin{figure*}[ht]
\centering
\includegraphics[width=80mm]{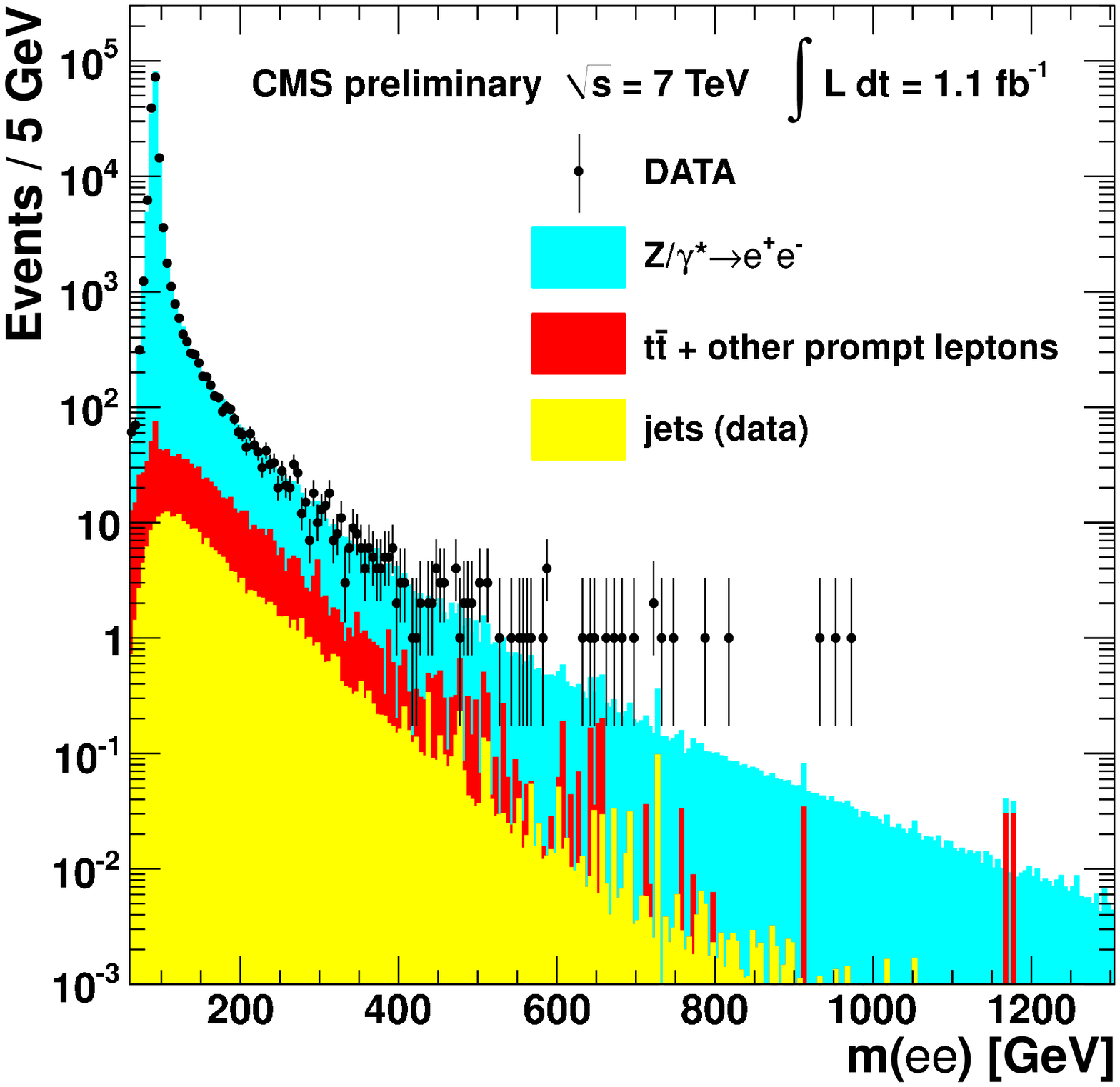}
\includegraphics[width=80mm]{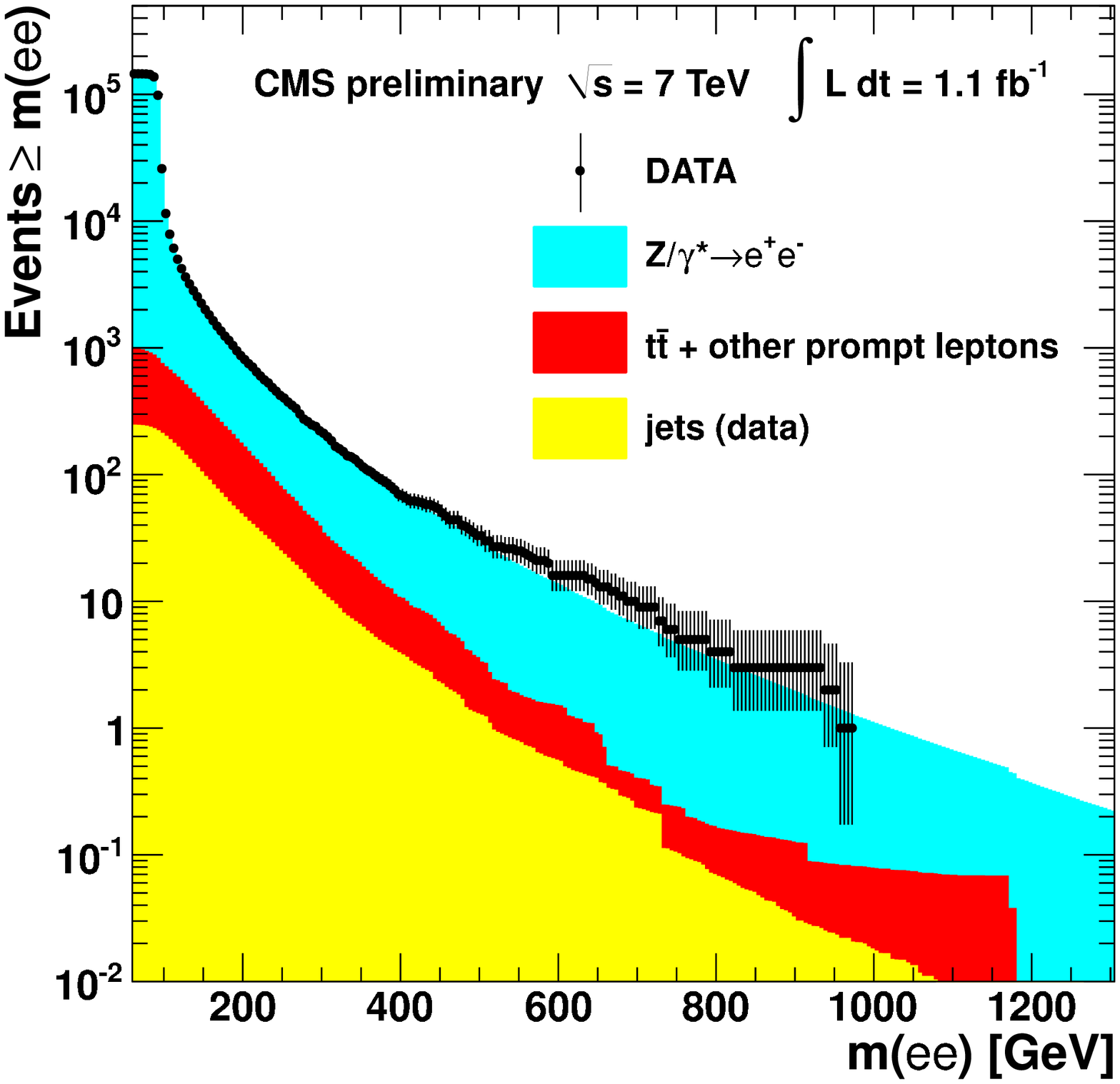}
\caption{Dielectron invariant mass (left) and the corresponding 
  cumulative spectrum (right).
  Individual components are normalized to NLO and then together
  to the Z-boson peak.
} 
\label{fig:dielectron_mass}
\end{figure*}

The Drell-Yan process produces the dominant irreducible
background, with the next biggest contribution from the
top pair and other top-like processes (tW, diboson and
$Z\to\tau\tau$). The remaining background
comes from jet misidentification as leptons ($1\%-5\%$ depending
on the channel), and from cosmic muons in the dimuon channel.
We found that the contribution from the latter, and 
from diphoton processes
misreconstructed as dielectrons are negligible.
Figures~\ref{fig:dimuon_mass} and \ref{fig:dielectron_mass}
depict the observed dilepton data overlaid with
the background components. The individual components
are normalized to next-to-leading order, and then
to the Z-boson peak in data.
The overall background rate and the shape of the dilepton
invariant mass distribution are taken from
the Drell-Yan Monte Carlo corrected to 
next-to-next-to-leading-order with 
\texttt{FEWZz v1.X} \cite{FEWZ},
\texttt{PYTHIA v6.409} and
\texttt{CTEQ6.1} PDF \cite{Stump:2003yu}.
For the purposes of setting the limits on the dilepton
resonance cross section,
the variation in the shape due to added
top-like and other background sources ($5\%-10\%$),
the uncertainties in k-factor, generator choice
and PDF sets are covered conservatively by
a background rate uncertainty of $20\%$($15\%$) in
the dimuon (dielectron) channel.

\begin{figure*}[ht]
\centering
\includegraphics[width=80mm]{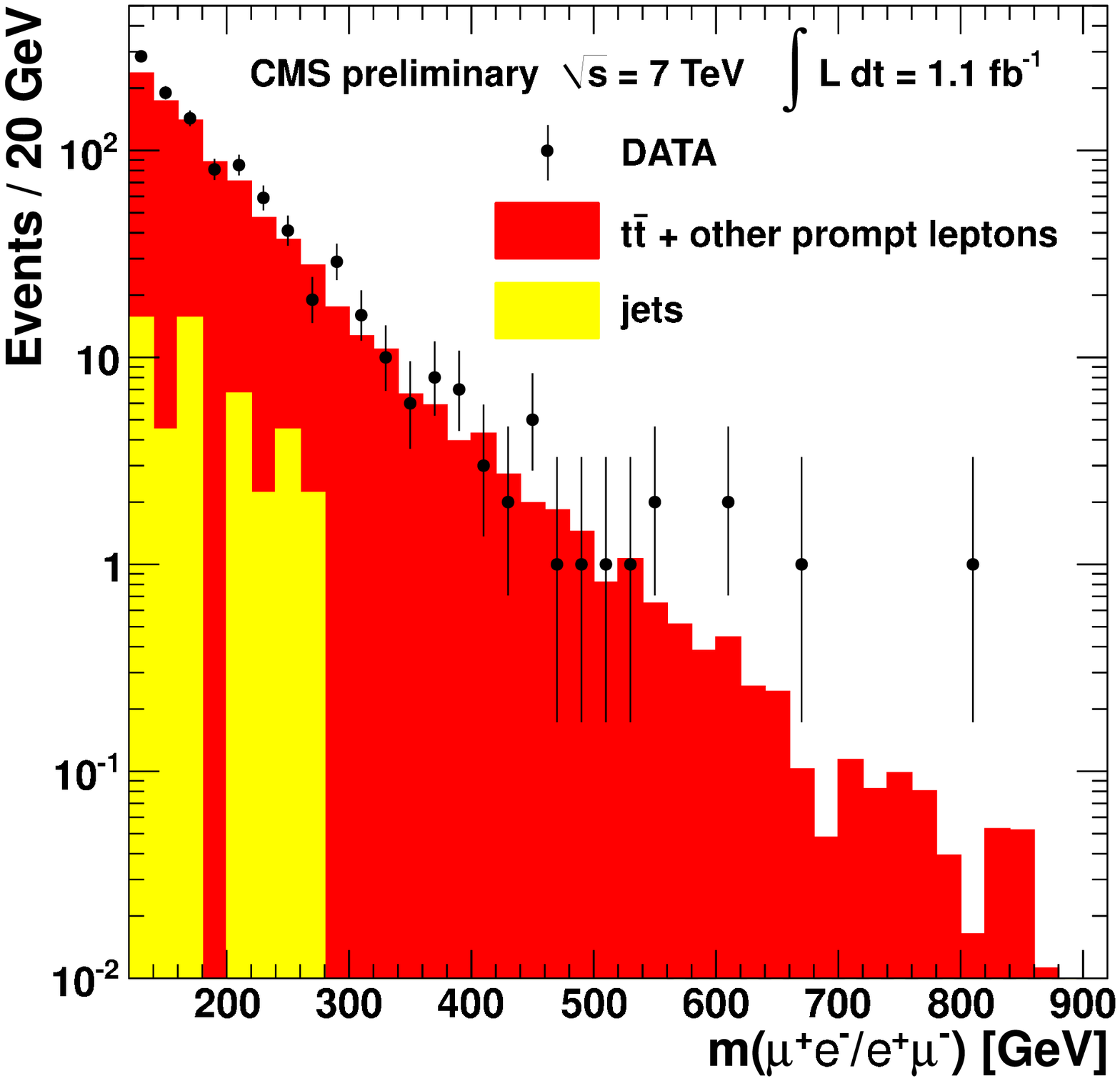}
\caption{Invariant mass of an electron and a muon
  of the opposite charge.} \label{fig:bg_top_emu}
\end{figure*}

As a cross check of the top-like background model,
we compare data and Monte Carlo distributions of the
dilepton invariant mass where the flavor and electric
charge of the two leptons are required to be 
different (``e$\mu$'' method). The reasoning is that if the
two leptons do not originate from a resonance, there is no special
reason for them to be of the same flavor. For each dielectron
and dimuon event, we expect to observe nearly two $e\mu$ 
events (the actual ratio is slightly different due to different
efficiencies for electrons and muons).
Figure~\ref{fig:bg_top_emu} demonstrates the comparison between data
and Mote Carlo for the $e\mu$ events, which we find
satisfactory.

\section{Statistical Inference}
We set $95\%$ C.L. upper limits on the cross section ratio
as defined in Equation~\ref{eq:poi}, assuming uniform prior
on the parameter of interest and Lognormal likelihood
constraint terms
on the nuisance parameters in order to model systematic
uncertainties.
We use the likelihood formalism to estimate the model parameters (via maximum likelihood, ML), 
and to build a likelihood ratio to be used as a test statistic. 
In the Bayesian methods the likelihood is further multiplied by 
priors to obtain the posterior pdf.
We define 
the unbinned likelihood for a data set as
\beq
\label{eq:likelihood}
\L(\bx | \btheta, \bnu) = \prod_{i=1}^{N}f(x_i | \btheta, \bnu),
\eeq
where the product is over the events in the data set \bx,
$f(x | \btheta, \bnu)$ is the probability density function of the observable $x$,
$x_i$ is the value of  the observable in the $i-$th event,
\btheta is a vector of the model parameters of interest,
\bnu is a vector of nuisance parameters.

It is often convenient and advantageous to define an extended likelihood
by adding the Poisson term. It provides the normalization of the data
in terms of the event yield:
\beq
\label{eq:extlikelihood}
\L(\bx | \mu, \btheta, \bnu) = \frac{\mu^N e^{-\mu}}{N!}\prod_{i=1}^{N}f(x_i | \btheta, \bnu),
\eeq
where
$N$ is the number of events in the data sample $\{x_i\}$,
$\mu$ is the Poisson mean number of events. In the 
following we will use extended likelihoods everywhere.

It is useful to define the profile likelihood ratio test statistic
\beq
\label{eq:llr}
t_\theta = -2\ln\llr(\btheta)=
-2\ln{\frac{\L_\mathtt{B}(\btheta, \hat{\hat{\bnu}}_\mathtt{B})}{\L_\mathtt{S+B}(\btheta, \hat{\hat{\bnu}}_\mathtt{SB})}},
\eeq
where $\L_\mathtt{B}$ and $\L_\mathtt{S+B}$ are the likelihood values
for the background-only and for the signal-plus-background models, 
and $\bnu_\mathtt{B}$ is a subset of $\bnu_\mathtt{SB}$.
The \emph{hat} notation ($\hat{\,\,\,}$)
symbolizes the ML estimator, i.e. $\hat{\btheta}$ is the
value of \btheta that maximizes the likelihood for a given model.
The double hat notation ($\hat{\hat{\,\,\,}}$)
stands for a conditional ML estimator for a given value of a parameter
of interest, i.e. $\hat{\hat{\bnu}}$ is the value of \bnu that
maximizes the likelihood for a given model, for a given
value of \btheta.

There are several \Zpr models that we consider in the analysis.
In the Sequential Standard Model (SSM), the \Zpr has the 
same couplings as the standard model $Z$ boson. The $\Psi$ model is 
based on an $E_6$ gauge symmetry. For the models overview see Z-Boson searches in
\cite{Nakamura:2010zzi}. We use the SSM model by default 
everywhere unless explicitly stated otherwise.

For the \Zpr search, we define the signal-plus-background likelihood as
\beq
\label{eq:model1}
\L_\mathtt{S+B}(\bm | \theta, \bnu) = 
\frac{\mu^N e^{-\mu}}{N!}\prod_{i=1}^{N}
\left(
  \frac{\mu_\mathtt{S}(\theta)}{\mu}f_\mathtt{S}(m_i|\bnu_\mathtt{S})+
  \frac{\mu_\mathtt{B}}{\mu}f_\mathtt{B}(m_i|\bnu_\mathtt{B})
\right),
\eeq
where
\bm is the dataset in which $m_i$ is the value of
the observable $m$ (the invariant mass of the lepton pair)
in $i$-th event,
$\theta$ denotes the parameter of interest, either the cross section
or the cross section ratio, as defined further,
\bnu is the vector of the nuisance parameters,
$f_\mathtt{S}$ and $f_\mathtt{B}$ are the PDFs for
the signal and the background
(specific shapes are defined later in the document),
$N$ is the total number of events observed,
$\mu_\mathtt{S}$ and $\mu_\mathtt{B}$ are
the expected signal and the background event yields, respectively,
and $\mu=\mu_\mathtt{S}+\mu_\mathtt{B}$ is the total 
number of events expected. Note that the expected
signal yield $\mu_\mathtt{S}$ is
a function of the parameter of interest.
The parameter of interest is the cross section ratio
\beq
R_\sigma = \frac{\xseczp}{\xsecz},
\label{eq:poi}
\eeq
where \xsecz is the cross section multiplied by the branching 
ratio for \ppzll.
Such an approach allows to exclude
the uncertainty on the integrated luminosity
from the measurement. 
In this case, we parameterize the expected signal yield as
\beq
\label{eq:musig}
\mu_\mathtt{S} = 
N_Z \cdot R_\sigma \cdot \frac{\selEff(\Zpr) \cdot \acc(\Zpr) 
\cdot }{\selEff(\Z) \cdot \acc(Z)}
\equiv
N_Z \cdot R_\sigma \cdot R_\epsilon \cdot R_{\acc}.
\eeq
Here $N_Z$ is the observed number of \Z events,
and $R_\epsilon \cdot R_{\acc}$ denotes
the fraction in Equation~\ref{eq:musig}.
The PDF, which represents the \Zpr signal model, is
\beq
\label{eq:sig}
f_\mathtt{S}(\mll|M,\Gamma,w) = \bw(\mll|M,\Gamma)\otimes\gaus(0, w),
\eeq
where
\mll is the invariant mass of the two leptons (the observable),
\bw stands for the resonant Breit-Wigner shape,
$\Gamma$ is its width,
$w$ is the width of the Gaussian resolution function.
For combining multiple analysis channels, the corresponding
likelihoods are multiplied together in order to build the combined
likelihood.

\section{Systematic uncertainty}
We combine all systematic uncertainties
into three components
that we treat independently: an uncertainty on signal sensitivity 
(includes uncertainties on signal and Z acceptances and 
efficiencies and on the Z event count), 
the background rate uncertainty, which is described in 
Section~\ref{sec:bg},
and the mass scale uncertainty in the dielectron
channel ($1\%$).

\section{Results}
\begin{figure*}[ht]
\centering
\includegraphics[width=80mm]{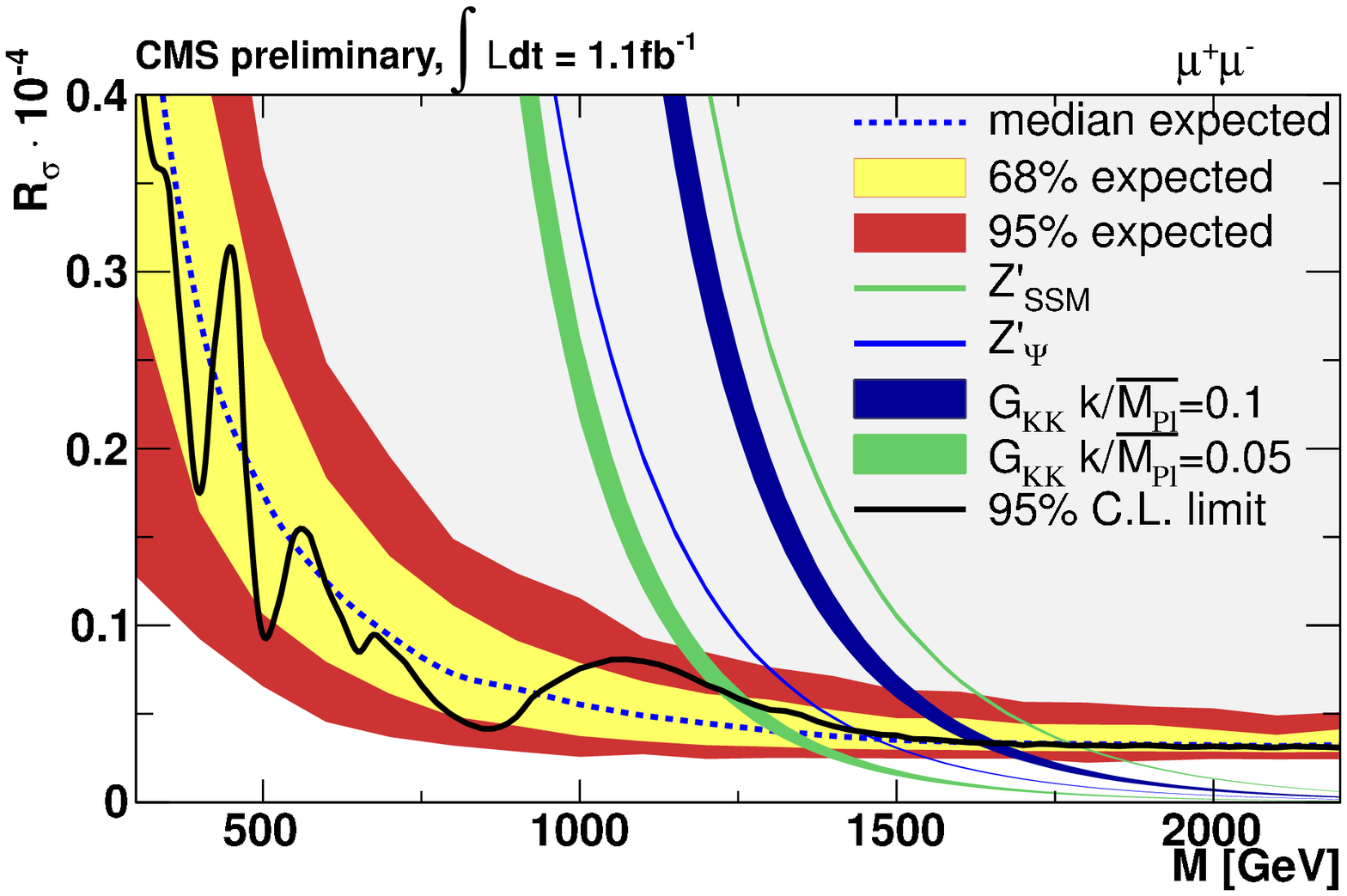}
\includegraphics[width=80mm]{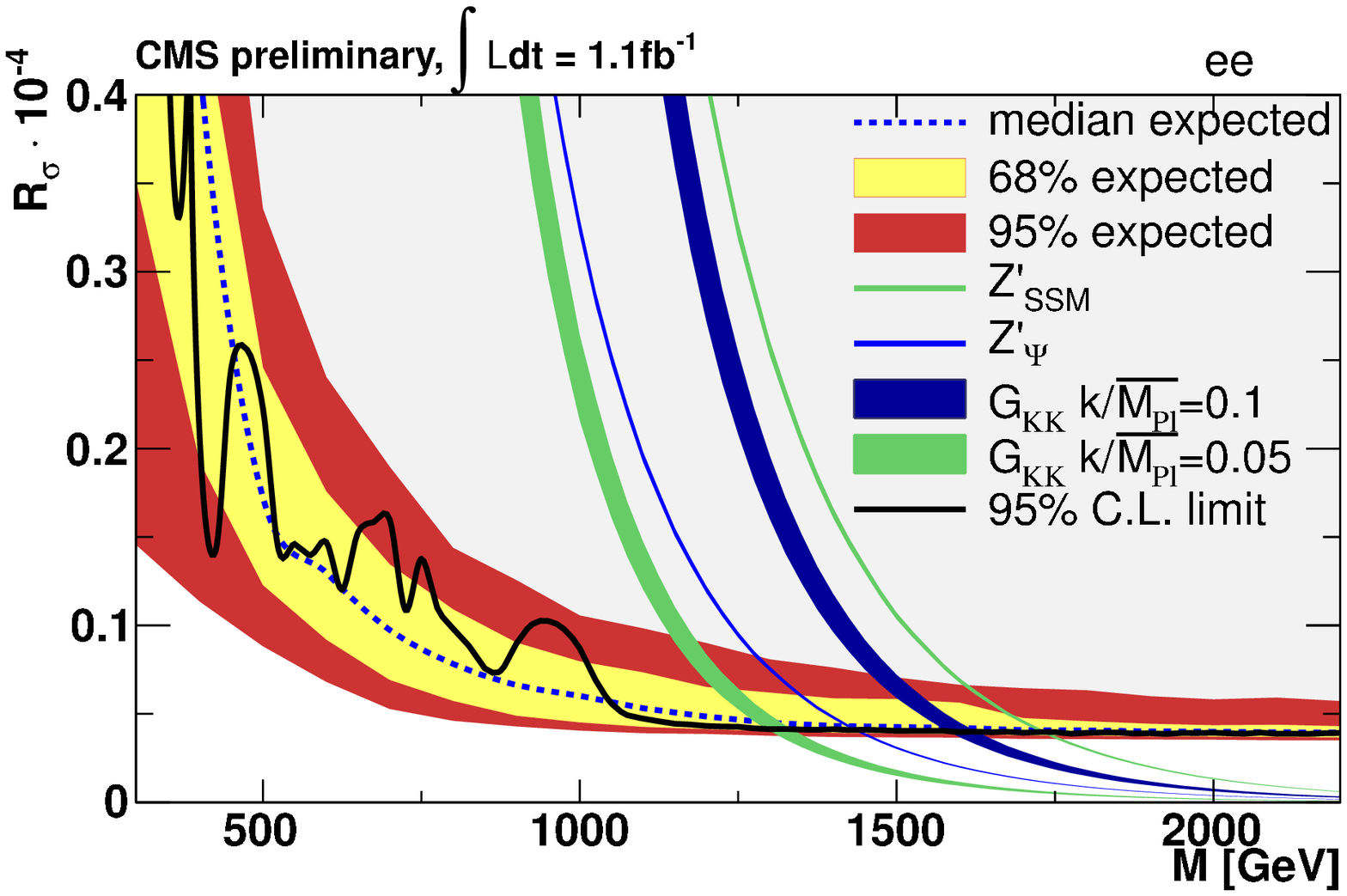}
\caption{Exclusion limits on the dilepton resonance
  cross section times branching fraction relative to
  the Z-boson standard model production, dimuon channel (left)
  and dielectron channel (right).
}
\label{fig:limit1}
\end{figure*}

\begin{figure*}[ht]
\centering
\includegraphics[width=135mm]{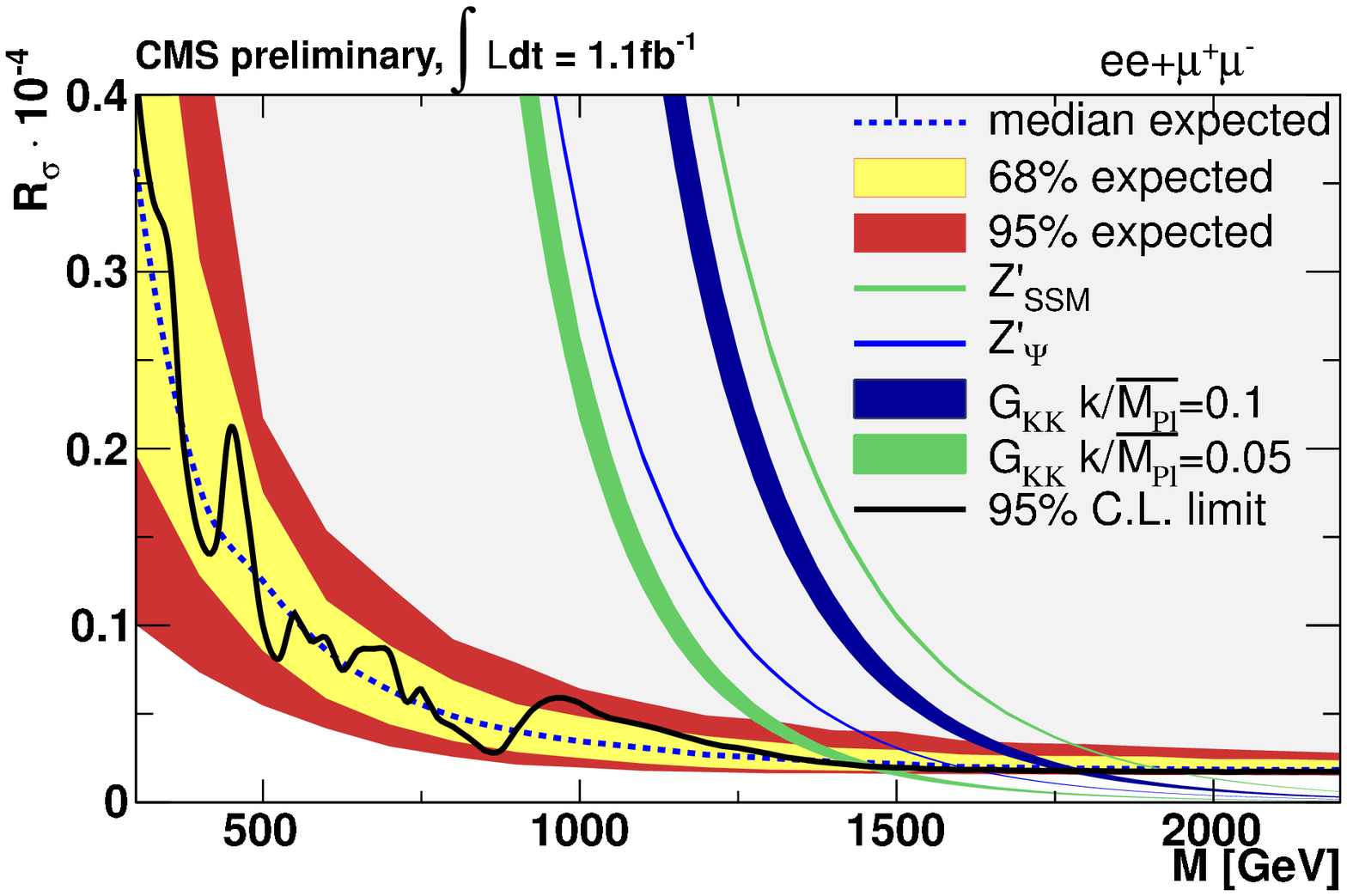}
\caption{Combined dimuon and dielectron channel exclusion 
  limits on the dilepton resonance
  cross section times branching fraction relative to
  the Z-boson standard model production.}
\label{fig:limit2}
\end{figure*}

We present reconstructed dilepton invariant mass distributions
in the CMS data in 
Figures~\ref{fig:dimuon_mass} and \ref{fig:dielectron_mass}
superimposed with the individual background components from
Monte Carlo. We use the mass spectra to set $95\%$ C.L.
Bayesian upper limits on the dilepton resonance cross section
ratio (Equation~\ref{eq:poi}). We use several popular theoretical
models to set lower limits on the corresponding resonance
masses, including the Sequential Standard Model $\Zpr$.
Figure~\ref{fig:limit1} displays the observed limits overlaid
with the median expected limits and 1- and 2-standard deviation
quantile bands. Theoretical estimates for four popular
theoretical models are overlaid as well.
Figure~\ref{fig:limit2} displays similar limit plots for the
combination of the dimuon and dielectron channels.
the likelihood ratio in Equation~\ref{eq:llr} is asymptotically distributed as a 
\chisq distribution with number of degrees of freedom equal to the difference 
in the numbers of free parameters between the two models.

By combining the two channels, we exclude \Zpr masses for 
the SSM and E6-motivated $\Psi$ model below 1940\gev and 1620\gev,
respectively. The corresponding limits in the individual
dimuon(dielectron) channels are 1780(1730)\gev and 1440(1440)\gev.
Combined analysis excludes masses of the RS Kaluza-Klein
gravitons for couplings of 0.05 and 0.10 at 1450\gev and 
1780\gev. The corresponding dimuon(dielectron) numbers are
1240(1300)\gev and 1640(1590)\gev.

\begin{figure*}[ht]
\centering
\includegraphics[width=80mm]{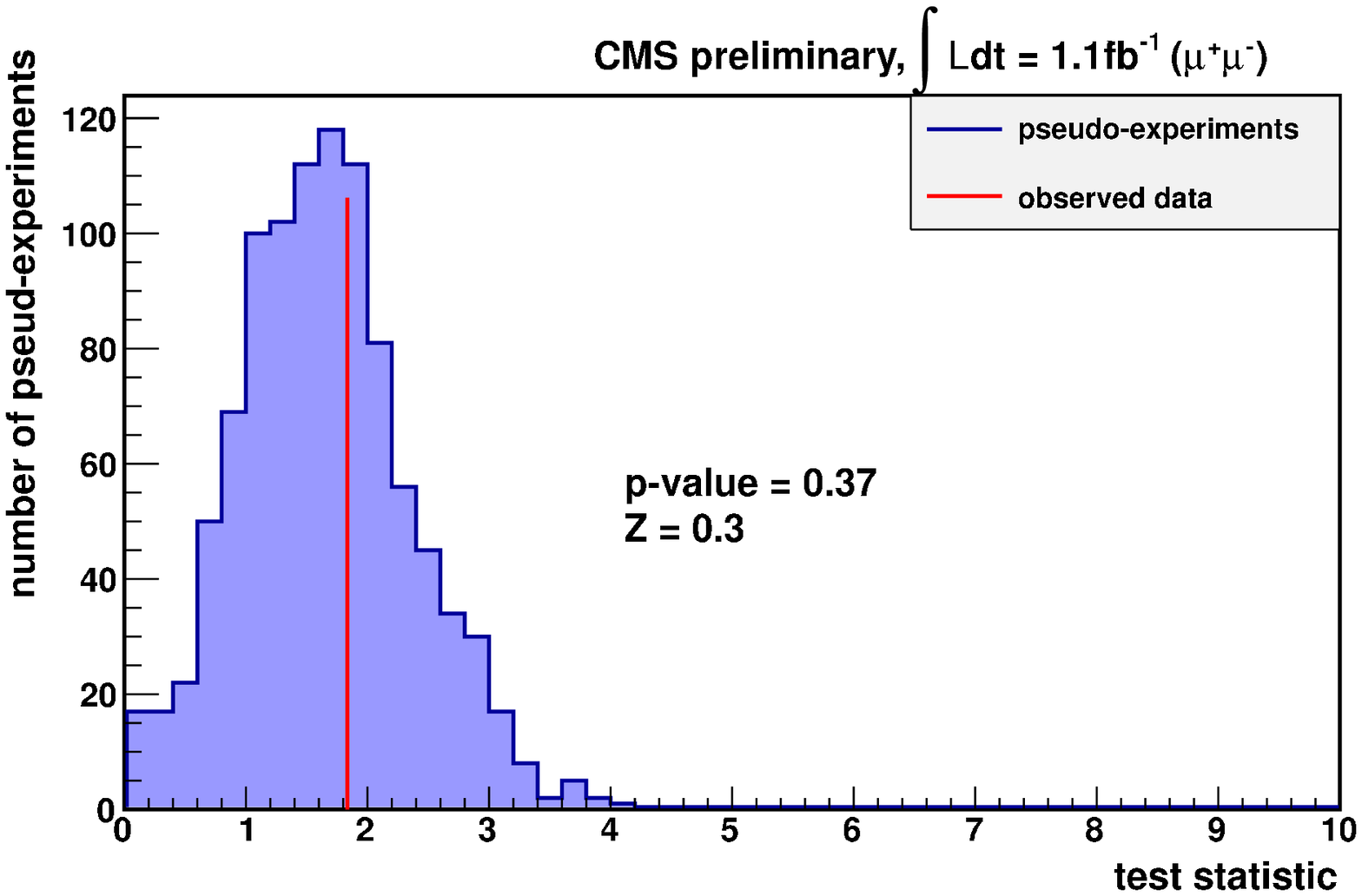}
\includegraphics[width=80mm]{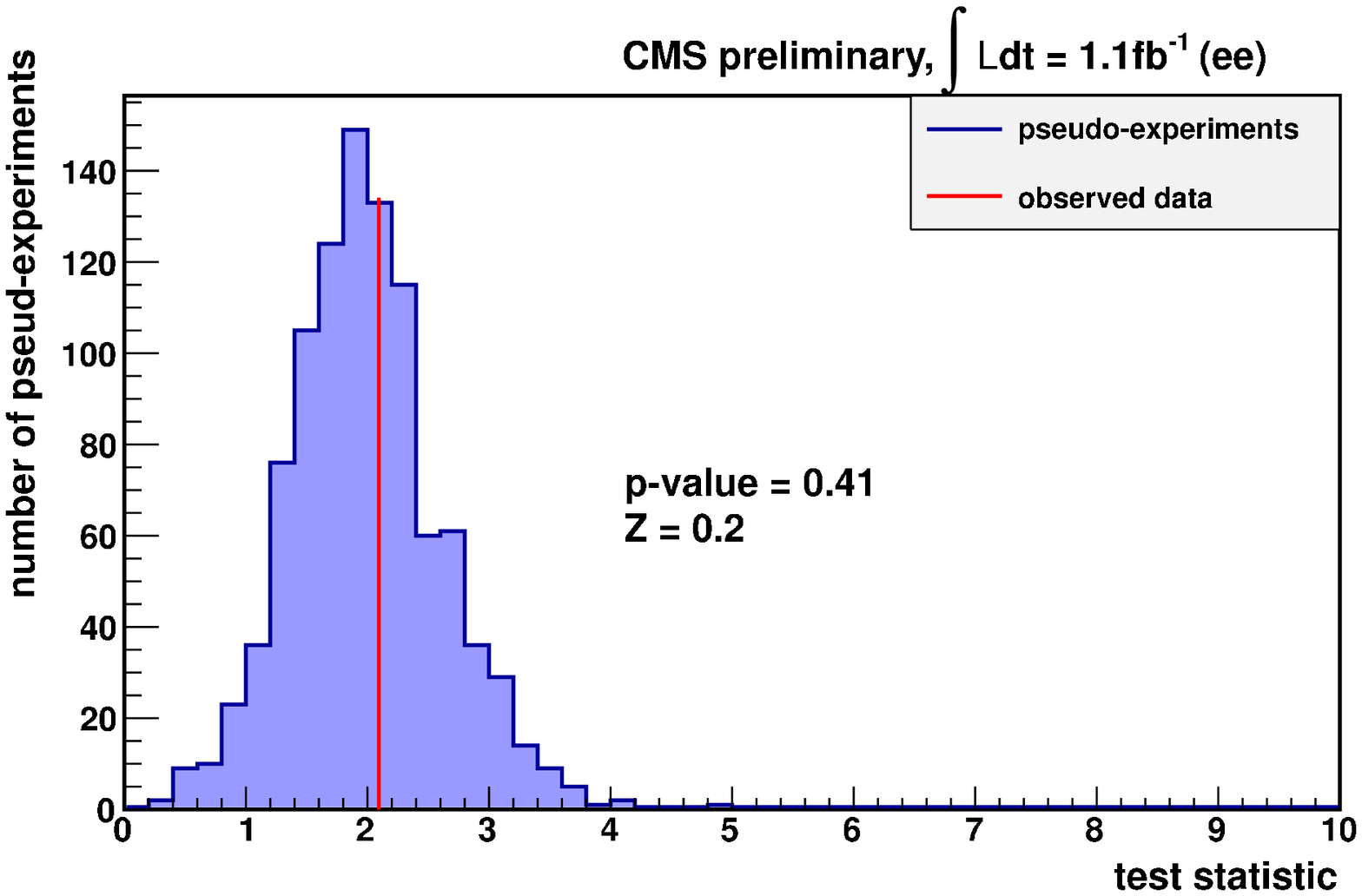}
\caption{Significance in the dimuon channel (left) and
  in the dielectron channel (right).
  A histogram corresponds to an ensemble of background-only
  pseudoexperiments. The red line is a value observed in
  data. A plotted value corresponds to the most signal-like
  pattern in a dataset, in a fine scan of the spectrum
  over the allowed invariant
  mass values.
}
\label{fig:lee1}
\end{figure*}

\begin{figure*}[ht]
\centering
\includegraphics[width=80mm]{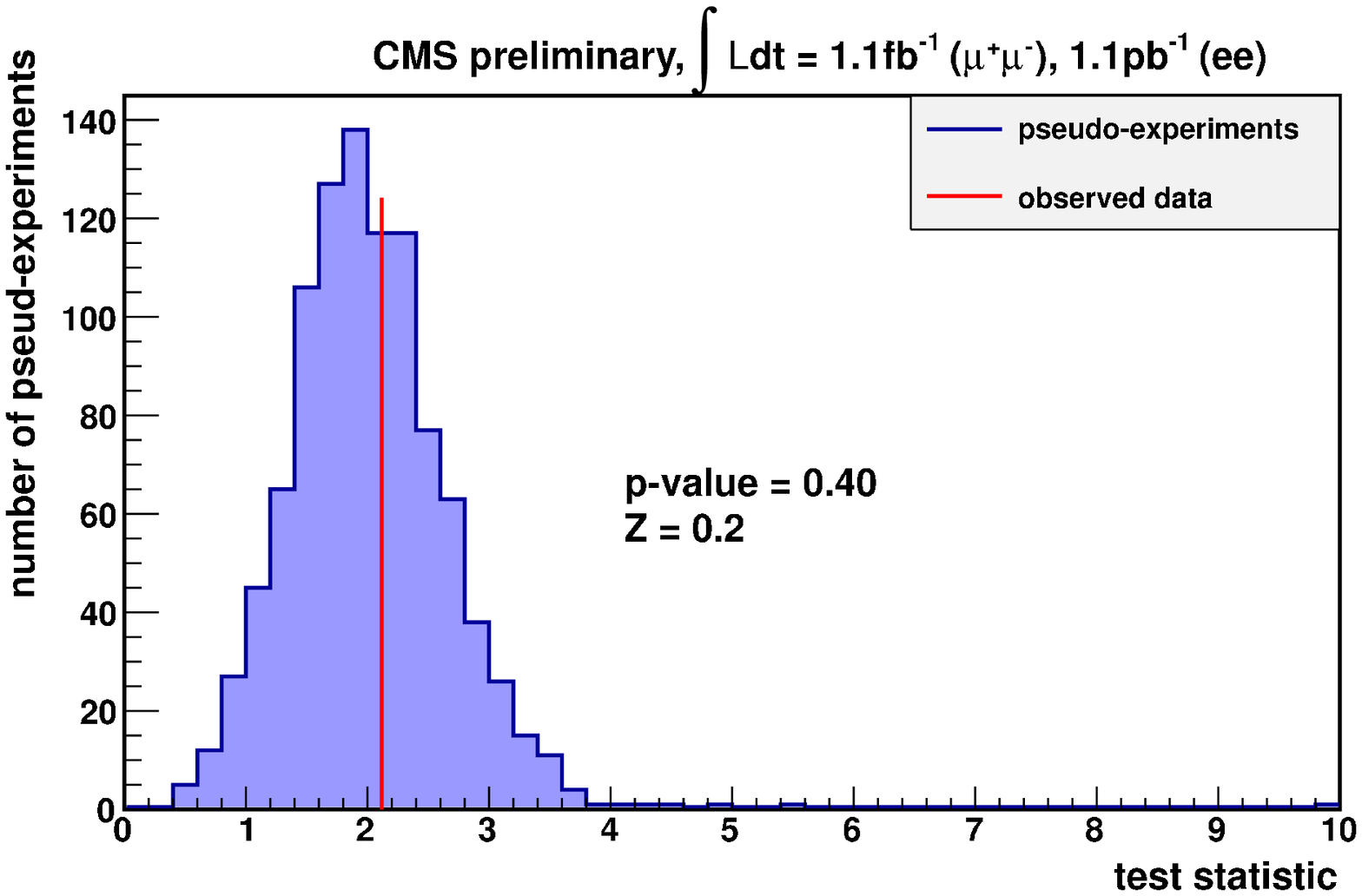}
\caption{Combined significance in the two channels.
  A histogram corresponds to an ensemble of background-only
  pseudoexperiments. The red line is a value observed in
  data. A plotted value corresponds to the most signal-like
  pattern in a dataset, in a fine scan of the spectrum
  over the allowed invariant
  mass values.
}
\label{fig:lee2}
\end{figure*}

We identify the most signal-like patterns in the data.
They correspond to a dimuon resonance at 1080\gev and
a dielectron resonance at 950\gev, with
local significances of 1.7 and 2.2 standard deviations,
respectively. Corrected for the ``trials factor'' (a consideration
that a signal-like fluctuation can happen at an arbitrary mass
value), the significances become 0.3 and 0.2 standard deviations,
respectively. Combined analysis suggest a dilepton resonance-like
signature at 970\gev with local significance of 2.1 and
significance corrected for the trials factor of 0.2 standard
deviations.
Figures~\ref{fig:lee1} and \ref{fig:lee2} display the sampling
distributions of the likelihood ratio test 
statistic (\ref{eq:llr}) obtained from ensembles of 
background-only pseudoexperiments, used for estimating
significances including the trials factor corrections,
overlaid with the value from data.

\section{Conclusion}
The CMS Collaboration has searched for high-mass narrow resonances in the
dilepton invariant mass spectra in the dimuon and 
the dielectron channels, using 1.1\invfb of integrated luminosity
recorded by the CMS detector operating at the LHC proton-proton
collider at CERN, with the center-of-mass energy of 7\tev.
The individual channel and combined spectra are consistent
with the Standard Model expectations. The 95\% C.L. upper limits
have been set on the product of the new resonance production
cross section and the corresponding branching fraction relative 
to the Standard Model Z boson production. Mass limits have been
set on the resonances predicted by the SSM and $\Psi$ \Zpr models,
and on the RS Kaluza-Klein gravitons for couplings of 0.05
and 0.1.

\begin{acknowledgments}
We wish to congratulate our colleagues in the CERN accelerator
departments for the excellent performance of the LHC machine. We thank
the technical and administrative staff at CERN and other CMS
institutes, and acknowledge support from: FMSR (Austria); FNRS and FWO
(Belgium); CNPq, CAPES, FAPERJ, and FAPESP (Brazil); MES (Bulgaria);
CERN; CAS, MoST, and NSFC (China); COLCIENCIAS (Colombia); MSES
(Croatia); RPF (Cyprus); Academy of Sciences and NICPB (Estonia);
Academy of Finland, ME, and HIP (Finland); CEA and CNRS/IN2P3
(France); BMBF, DFG, and HGF (Germany); GSRT (Greece); OTKA and NKTH
(Hungary); DAE and DST (India); IPM (Iran); SFI (Ireland); INFN
(Italy); NRF and WCU (Korea); LAS (Lithuania); CINVESTAV, CONACYT,
SEP, and UASLP-FAI (Mexico); PAEC (Pakistan); SCSR (Poland); FCT
(Portugal); JINR (Armenia, Belarus, Georgia, Ukraine, Uzbekistan); MST
and MAE (Russia); MSTD (Serbia); MICINN and CPAN (Spain); Swiss
Funding Agencies (Switzerland); NSC (Taipei); TUBITAK and TAEK
(Turkey); STFC (United Kingdom); DOE and NSF (USA).
\end{acknowledgments}

\bigskip
\bibliographystyle{Science}
\bibliography{kukartsev_zprime_dpf2011}

\end{document}